\documentclass[format=acmsmall, review=false, screen=false]{acmart}

\usepackage{booktabs} 

\usepackage[font=small,labelfont=bf]{caption}
\usepackage{multicol}
\usepackage{amsmath}
\usepackage{algorithmic}
\usepackage{algorithm}
\usepackage{array}
\usepackage{makecell}
\usepackage{caption}
\usepackage{adjustbox}

\setcopyright{none}

\begin{document}

\title[Edge Cloud Offloading Algorithms]{Edge Cloud Offloading Algorithms: Issues, Methods, and Perspectives}  
\author{Jianyu Wang}
\affiliation{%
  \institution{University of Missouri-St. Louis}
  \department{Department of Mathematics and Computer Science}
  \city{St. Louis}
  \state{MO}
  \postcode{63121}
  \country{USA}}
  
\author{Jianli Pan}
\affiliation{%
  \institution{University of Missouri-St. Louis}
  \department{Department of Mathematics and Computer Science}
  \city{St. Louis}
  \state{MO}
  \postcode{63121}
  \country{USA}}

\author{Flavio Esposito}
\affiliation{%
  \institution{Saint Louis University}
   \city{St. Louis}
  \state{MO}
  \country{USA}}
  
\author{Prasad Calyam}
\affiliation{%
  \institution{University of Missouri-Columbia}
   \city{Columbia}
  \state{MO}
  \country{USA}}
  
\author{Zhicheng Yang}
\affiliation{%
  \institution{University of California, Davis}
  \state{CA}
  \country{USA}}
  
\author{Prasant Mohapatra}
\affiliation{%
  \institution{University of California, Davis}
  \state{CA}
  \country{USA}}

\begin{abstract}
Mobile devices supporting the "Internet of Things" (IoT), often have limited capabilities in computation, battery energy, and storage space, especially to support resource-intensive applications involving virtual reality (VR), augmented reality (AR), multimedia delivery and artificial intelligence (AI), which could require broad bandwidth, low response latency and large computational power. Edge cloud or edge computing is an emerging topic and technology that can tackle the deficiency of the currently centralized-only cloud computing model and move the computation and storage resource closer to the devices in support of the above-mentioned applications. To make this happen, efficient coordination mechanisms and ``offloading'' algorithms are needed to allow the mobile devices and the edge cloud to work together smoothly. In this survey paper, we investigate the key issues, methods, and various state-of-the-art efforts related to the offloading problem. We adopt a new characterizing model to study the whole process of offloading from mobile devices to the edge cloud. Through comprehensive discussions, we aim to draw an overall ``big picture'' on the existing efforts and research directions. Our study also indicates that the offloading algorithms in edge cloud have demonstrated profound potentials for future technology and application development. 
\end{abstract}

%
%


%
%


\thanks{First manuscript: April 30th, 2017.

Author's addresses: J. Wang {and} J. Pan , the Department of Mathematics and Computer Science, University of Missouri-St. Louis, MO 63121 USA; email: jwgxc@umsl.edu, pan@umsl.edu;
F. Esposito, the Department of Computer Science, Saint Louis University, MO 63103 USA.; 
P. Calyam, the Department of Computer Science, University of Missouri-Columbia, MO 65211 USA; 
Z. Yang {and} P. Mohapatra, the Department of Computer Science, University of California, Davis, CA 95616 USA; 
}

\maketitle
\renewcommand{\shortauthors}{J. Wang et al.}

\section{Introduction}
\label{sec:intro}

We are embracing the future world of 5G communication and Internet of Things (IoT). Smart mobile devices are becoming more popular and playing increasingly important roles in every aspect of our daily life~\cite{Book1}. Various applications running on these mobile devices require not only bounded latency, wide bandwidth and high computation performance, but also long battery life~\cite{Chang1}. In these cases, mobile devices alone are insufficient as they suffer from limited local capabilities of computation and energy to deliver performant resources-intensive applications. Therefore, such resource gap is fulfilled by the remote and centralized data centers or clouds services such as Amazon Web Services~\cite{AWS}, Microsoft Azure~\cite{Microsoft} and Google Cloud~\cite{Google}. These centralized clouds (CCs) can offer virtually unlimited computation, networking and storage resources. For many years, the cloud elasticity model has been widely successful and an added value for both enterprises and cloud providers. However, recent advances in resource intensive IoT applications such as face recognition, ultra-high-definition video, augmented reality (AR), virtual reality (VR) and voice semantic analysis, are challenging the scalability and resiliency models of the traditional cloud computing with more rigorous demands in response latency and data storage. Moreover, the current limited bandwidth of the backbone network cannot afford the back-and-forth transmission of the exponentially increasing amount of data generated by the future IoT devices for the ever-increasing mobile applications. 

To tackle the above challenges, Edge Clouds (ECs) (also called ``fog computing''~\cite{bonomi2012fog} or ``Mobile Edge Computing''~\cite{2016mobileedge} in some literatures) have been proposed. The core idea is to add resources at the network edge; in particular computation, bandwidth, and storage resource are moved closer to the IoT devices to reduce the backbone data traffic and the response latency, and to facilitate the resource-intensive IoT applications. EC has relatively smaller computation capacity compared to CC, but takes advantage of short access distance, flexible geographical distribution, and relatively richer computational resource than mobile devices. 




Other than the common mobile applications, ECs are also more suitable for some special circumstances. For example, in disaster or battlefield environments, ECs can be very useful in providing uninterrupted communication and handling intensive parallel tasks with high accuracy and low communication latency even if the backbone network access is not available. The second example is in health monitoring or remote access to medical devices~\cite{MedicalDevice1}~\cite{MedicalDevice2}, where patients are equipped with numbers of wearable sensors to monitor vital signs in real time. ECs could process the data collected from these sensors to extend their battery life and generate quick response in emergency to save lives. Finally, with the advent of IoT, ECs can play a key role to build a fundamental tier of IoT systems for many more distributed applications in smart home, smart health, smart vehicles and even smart cities~\cite{Book1}. Recent research such as HomeCloud framework~\cite{HomeCloud} and Incident-Supporting Visual Cloud~\cite{IncidentCloud} concentrate on the combinative applications between EC and IoT.

\begin{figure}
\centering
\includegraphics[width=5.1in, height=1.6in]{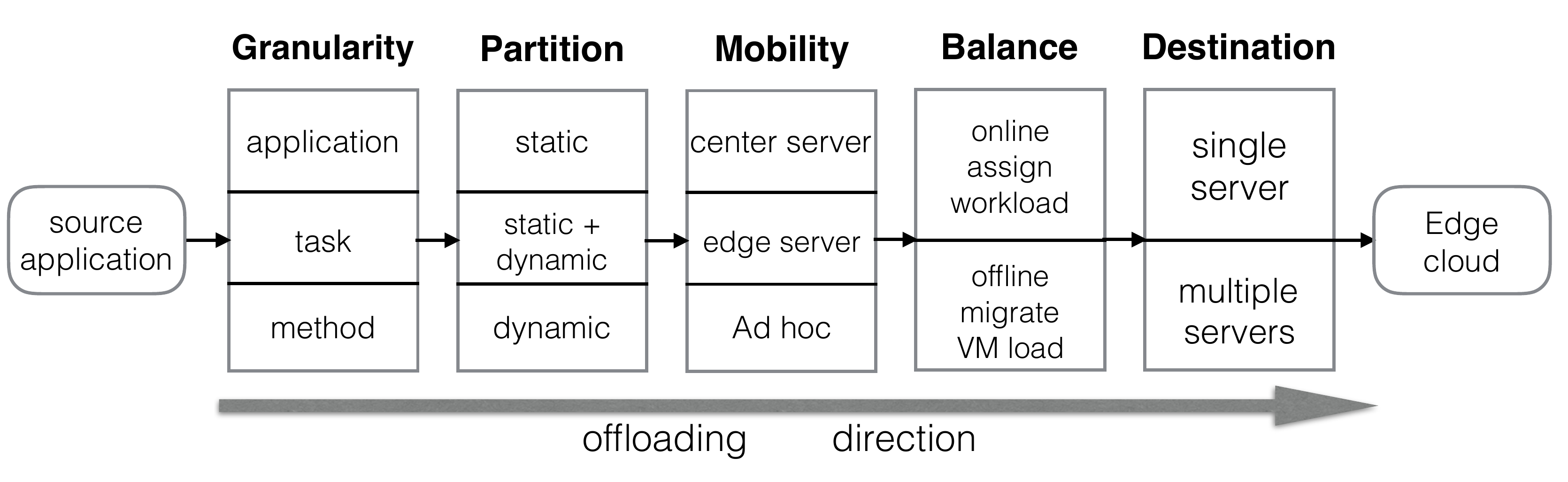}
\captionsetup{justification=centering}
\caption{The whole process of offloading from mobile devices to an edge cloud.}
\label{fig:offloading_process}
\end{figure}


In such an edge cloud vision, the EC offloading problem, i.e., the problem of transmitting a workload from a mobile device to the ECs is one of the principal challenges. Offloading algorithms are of central importance for an efficient coordination between the ECs and the mobile devices. Our paper surveys the recent representative offloading algorithms. In particular, we adopt a novel characterizing model that serves as taxonomy to study process of offloading from mobile devices to the ECs. The process responsibility is divided among three main agents: mobile devices, communication links and ECs. Specifically, mobile devices are responsible for determining how an application is partitioned, which parts should be executed locally or remotely, and the offloading scheme. The communication link is influenced by fluctuation of bandwidth, connectivity and device mobility. EC servers handle the balance of server load to achieve maximum service rates and system throughput. 

Given this offloading model, we classify existing solutions using five dimensions: \textit{offloading destination}, \textit{EC load balancing}, \textit{user devices mobility}, \textit{application partitioning} and \textit{partition granularity} as illustrated in Fig.~\ref{fig:offloading_process} from right to left. Such classification covers the whole offloading process in a sequential order from mobile devices to ECs. By analyzing the problem definition, mathematical models and optimization solutions, each algorithm discussed in this paper stands for a typical and creative research direction. 


Several other survey papers related to edge cloud	\cite{survey1}\cite{survey2}\cite{survey3} discuss edge cloud computing in different domains (such as edge computing architecture, communication, computation offloading and use case studies).  However, our paper is different with them in collecting offloading algorithms and analyze their mathematical models from a holistic comprehensive perspective.


The rest of this paper is organized as follow. Section \ref{sec:destination} introduces scenarios of the single and multiple servers as offloading destination. Section \ref{sec:balance} shows the online and offloading methods to dynamically balance server load. Section \ref{sec:mobility} analyzes scenarios in which mobile devices lose connectivity due to their continuous movement and the corresponding solutions. Section \ref{sec:offloadingPartition} presents the schemes to offload partitioned components according to its internal execution order and cost. Section \ref{sec:granularity} reviews the granularity of application partitioning and explains their advantages and disadvantages. Section \ref{sec:discussions} discusses the related mathematical models, future challenges as well as technology trends. Finally, the conclusions follow in section \ref{sec:conclusion}.

\section{Offloading Destination - single server vs. multiple servers} \label{sec:destination} 
\begin{figure}
\centering
\includegraphics[width=3.1in, height=2.2in]{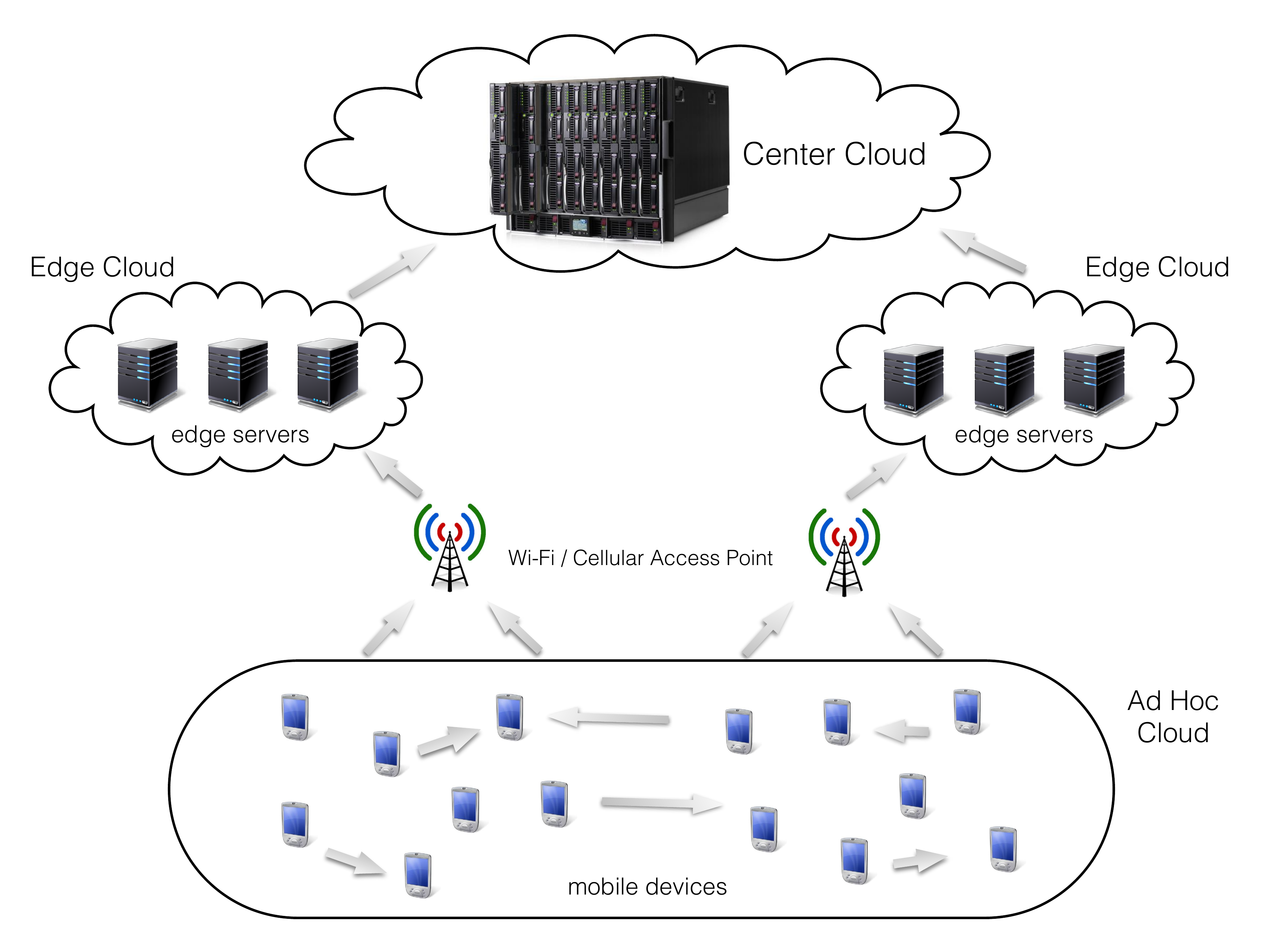}
\captionsetup{justification=centering}
\caption{Mobile cloud computing framework.}
\label{fig:mobile_framework}
\end{figure}

Edge offloading is a strategy to transfer computations from the resource-limited mobile device to resource-rich cloud nodes in order to improve the execution performance of mobile applications. The selection of cloud servers is worth careful consideration at the beginning of the design phase of an offloading algorithm. The workload of mobile devices at the run time could be offloaded to only one server for sequential execution or to multiple servers for parallel execution, leading to a lower response latency. Fig.~\ref{fig:mobile_framework} shows the framework of mobile cloud computing that illustrates communication relationship among all functional components. User devices are distributedly located at the edge of the network. They could offload computation to EC servers via WiFi or cellular networks. If a single EC server cannot is insufficient to sustain the workload in peak periods, other EC servers or CC servers are made available to assist the application. Current studies on offloading focus on a wide range of applications, where the requirements of network and computing resources vary in different execution environments. For example, processes whose execution follow a sequential order are suitable to be offloaded to one single server since the communication among serialized parts is frequent. On the other hand, applications with repetitive computation are more suitable for parallelized servers. In this section, we discuss some representative algorithms based on such offloading destination taxonomy dimension.

\subsection{Single Server}
In this section we discuss MAUI~\cite{MAUI} and CloneCloud~\cite{CloneCloud} two algorithms that utilize a single-sever offloading strategy.


\smallskip

\subsubsection{MAUI}

MAUI is a fine-grained approach that offloads parts of programs remotely to solve the mobile energy problem~\cite{MAUI}. The remote server could be a CC server or a nearby EC server at WiFi access point. As a pioneer of all offloading systems, the MAUI offloading strategy takes advantage of program partitioning and full process migration, which also reduces developers' programming burden.

\begin{figure}
\centering
\includegraphics[width=2.6in, height=1.7in]{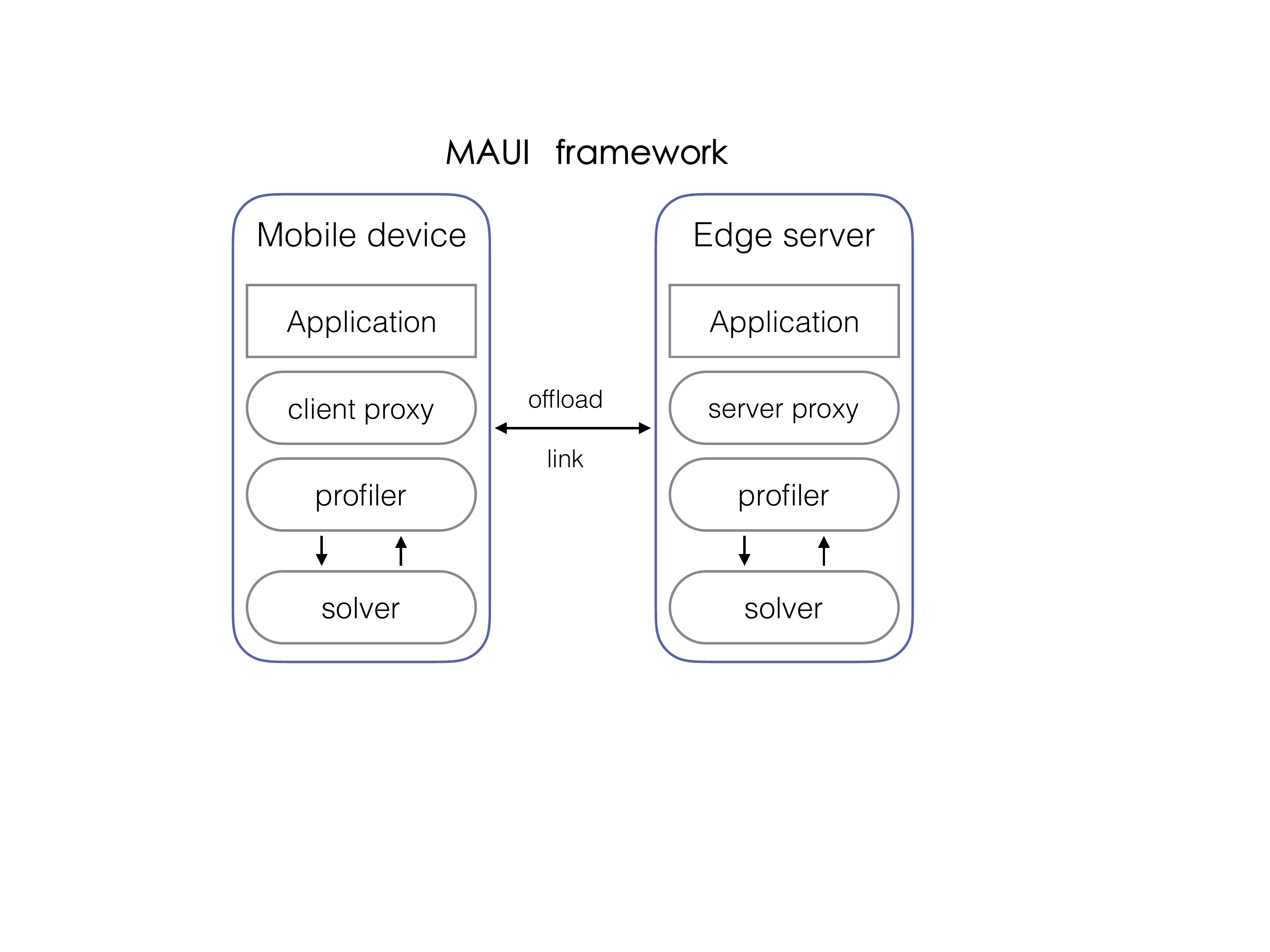}
\captionsetup{justification=centering}
\caption{MAUI system model.}
\label{fig:maui_model}
\end{figure}

The architecture of MAUI follows a client-server model. Both the server and mobile devices have three functional components: proxy, profiler and solver, as illustrated in Fig.~\ref{fig:maui_model}. The proxy is used to transmit data and control instructions. The profiler retrieves the data about program requirement, execution energy cost and network environment, while the solver decides the program partitioning strategy. Under such an architecture, MAUI models communication cost and computation cost with a 0-1 integer linear optimization problem which is derived from the method called graph. The graph is a flow diagram that presents the computation cycles, energy cost and data size at each stage of the application execution. When a method is called and a remote server is available, the optimization framework dynamically determines whether the method should be offloaded to maximize the total energy saving or not. The problem is defined as follow:
$$\sum_{v\in V}I_{v}\times E_{v}^{l} -  \sum_{(u,v)\in E} |I_{u}-I_{v}|\times C_{u,v}$$
constraints: $$\sum_{v\in V}((1-I_{v})\times T_{v}^{l} +(I_{v}\times T_{v}^{r})) + \sum_{(u,v)\in E} (|I_{u}-I_{v}|\times B_{u,v})\leq L$$
where $E_{v}^{l}$ is energy cost by executing method locally, $C_{u,v}$ is energy cost of tranferring data between nodes, $B_{u,v}$ is time of transferring data, $L$ is time latency limit, $l$ is locally, $r$ is remotely, $v, u$ are vertices on the call graph, $I_{v}$ is 0-1 choice.

Specifically, MAUI first helps program developers simply mark the methods as remotable that could be considered offloading to a server, then MAUI automatically decides which methods should be offloaded with the aid of programming reflecting and type-safety to manage program behavior. Second, at run time, the profiler start to periodically collect system information from three factors: device, program and network. Third, based on the factors information, a linear program solver uses data input by profilers to find a global optimization way for offloading. After obtaining offloading strategy, the proxies on the mobile devices and the server start to implement the solution, while the mobile application performs simultaneously. Proxies handle the exchange of priority to execute code locally or remotely. When the program on the server calls a method, which is assigned to a local device, the server transfers the execution control to the mobile device, and vice versa. Therefore, the synchronization between local and remote must work in serialization even if there is corresponding data transmission overhead.

Besides the system framework and the offloading algorithm, the authors in MAUI also investigated several approaches to evaluate the system's macro and micro performance. The macro-benchmarks contain energy consumption, performance of mobile applications and ability of supporting resource-intensive applications. The micro-benchmarks evaluate the overhead of each system components, the adjustment of each algorithm parameters, the change of network environment and the CPU costs. To test the effectiveness of MAUI, three kinds of popular applications on mobile device are being experimented:  resource-intensive face recognition, latency-sensitive video game, and voice language translation. With their implementation of MAUI, authors demonstrate a 27\% energy of the smartphone is saved for video game, 45\% for chess game and even more than 85\% for face recognition. Meanwhile, the authors also show how the latency improves by a factor of 4.8 times using nearby edge cloud servers.

\smallskip
\subsubsection{CloneCloud}

Similar to MAUI, CloneCloud is also a fine-grained approach that automatically identifies the communication and computation costs for migrating the workload from local to edge cloud~\cite{CloneCloud}. However, CloneCloud does not require any developer efforts in marking whether a part of the program can be offloaded or not. The application is unmodified and the suitable portions of its execution are offloaded to the edge end. CloneCloud owns a flexible application partitioner and executes the workload from mobile devices on the task-level virtual machines  (VM) such as Java virtual machine and .NET.

To establish a CloneCloud framework, a static analyzer, as the first step of partitioning mechanism, is used to discover constraints of the application to be executed on edge servers. The analyzer then determines the legal executable parts that qualify this set of constraints. If a code segment performs expensive processing and satisfies the constraints, it runs on the cloud server. There are three kinds of constraints for the analyzer. First, the chosen methods should not use the specific features which are pinned to the mobile machines such as GPS and various sensors. The methods of these features are natively embedded with hardware. Second, the methods that shared native state should be allocated to the same VM when they serve for the same application process. Third, the caller-caller relation among methods is monitored. If a partition point is located in the caller, it should not be in the callee. Such nested migration is prohibited to avoid triggerings multiple times the same migration at the same partition points.

The dynamic profilers collect the necessary data to build cost models based on the outputs of the analyzer when the various applications adapt different execution configuration. The cost model of an execution trace of an application is presented as a profile tree data structure, where nodes represent methods and edges represent cost values, such as execution time and energy consumption. Next, a mathematical optimization solver determines the migration points where the workload is offloaded at run time to minimize the overall cost of the mobile devices. 

Finally, the chosen program methods are offloaded to an available server with a clone VM installed. When the execution process on the mobile device reaches to the migration points, its process is suspended and its current state is packaged and transmitted. The clone VM will initiate a new thread with the packaged state in the stack and heap objects. Then the application process resumes on the cloud server. When the assigned tasks are completed, application state is repackaged and shipped back to the original mobile device, then the process on mobile device resumes.

With the help of the above offloading framework, CloneCloud achieves application partitioning and seamless cooperation between the local and the remote virtual processing instance. Experiments on tasks such as virus scanning and image search show that the algorithm helps applications achieve up to 20 times speedup and 20-fold decrease in energy consumption.

Aside from the advantages brought by offloading strategies, a set of new challenges appears. One is the inability to easily offload the workload caused by native functional modules such as camera, GPS and sensors. Second, when the offloading strategies try to permit perfunctory concurrency between unoffloaded and offloaded workload, the synchronization of application data should be prudently considered to keep data updated.

\smallskip
\subsection{Multiple Servers}

With the prosperity of computation intensive applications, we are facing more serious challenges on the energy and latency-sensitive for applications such as multimedia, 3D modeling of disaster site and unmanned driving. In these cases, the tolerance of execution latency is relatively rigorous to meet customer requirements. In these cases, and in others where the execution latency constraints are severe, a single server or VM may be unable to provide sufficient communication bandwidth and computing capability. To this aim, solutions featuring parallel offloading on multiple servers have been proposed to distribute partitioned tasks to a cluster of servers which have different capacities of computation and communication resources. In the next subsections we discuss a few representative solutions that adopt this model. 

\smallskip

\subsubsection{ThinkAir}
After the publication of MAUI and CloneCloud, ThinkAir ~\cite{ThinkAir}was proposed to cope with the disadvantages of these two systems; in particular, ThinkAir extends in new ways for resource allocation and parallel task execution. 



ThinkAir attains scalability by providing VMs which runs the same smartphone execution environment on the edge nodes synchronously. Moreover, the system improves the performance compared to CloneCloud by dynamically allocating cloud resources instead of statically analyzing partitions of applications. ThinkAir achieves such scalability and flexibility through two perspectives. First, parallel execution on several edge servers can satisfy the high computation requirements of the mobile applications such as face recognization. The system can divide a calculation problem into sub-problems to execute on the multiple VMs to reduce the waiting interval of the tasks between cloud and mobile devices. Seconds, the tolerance of energy consumption and latency fluctuates due to the resouce types of applications, the hardware performance of mobile devices, the limited battery capacity and the specific user configurations. Besides the above two advantages, ThinkAir also deals with the unstable connectivity of cloud service to guarantee the precise execution of applications.

Similar to MAUI, edge cloud application developers using ThinkAir would still need to modify the code for running application smoothly. However, such modification workload requirement seems less than the one required in MAUI's because ThinkAir provides programmers with a customized API and a compiler. Moreover, ThinkAir provides an execution controller that determines the necessity to offload a program method. When the method is executed for the first time, the decision is only based on the environmental parameters. The subsequent execution decision is determined by the combination of data including latency and energy cost in the past invocation and environmental data. There are execution controllers both on the user devices and cloud servers to determine the offloading node according to the requirement of execution time and energy.

\begin{figure}
\centering
\includegraphics[width=2.1in, height=2.5in]{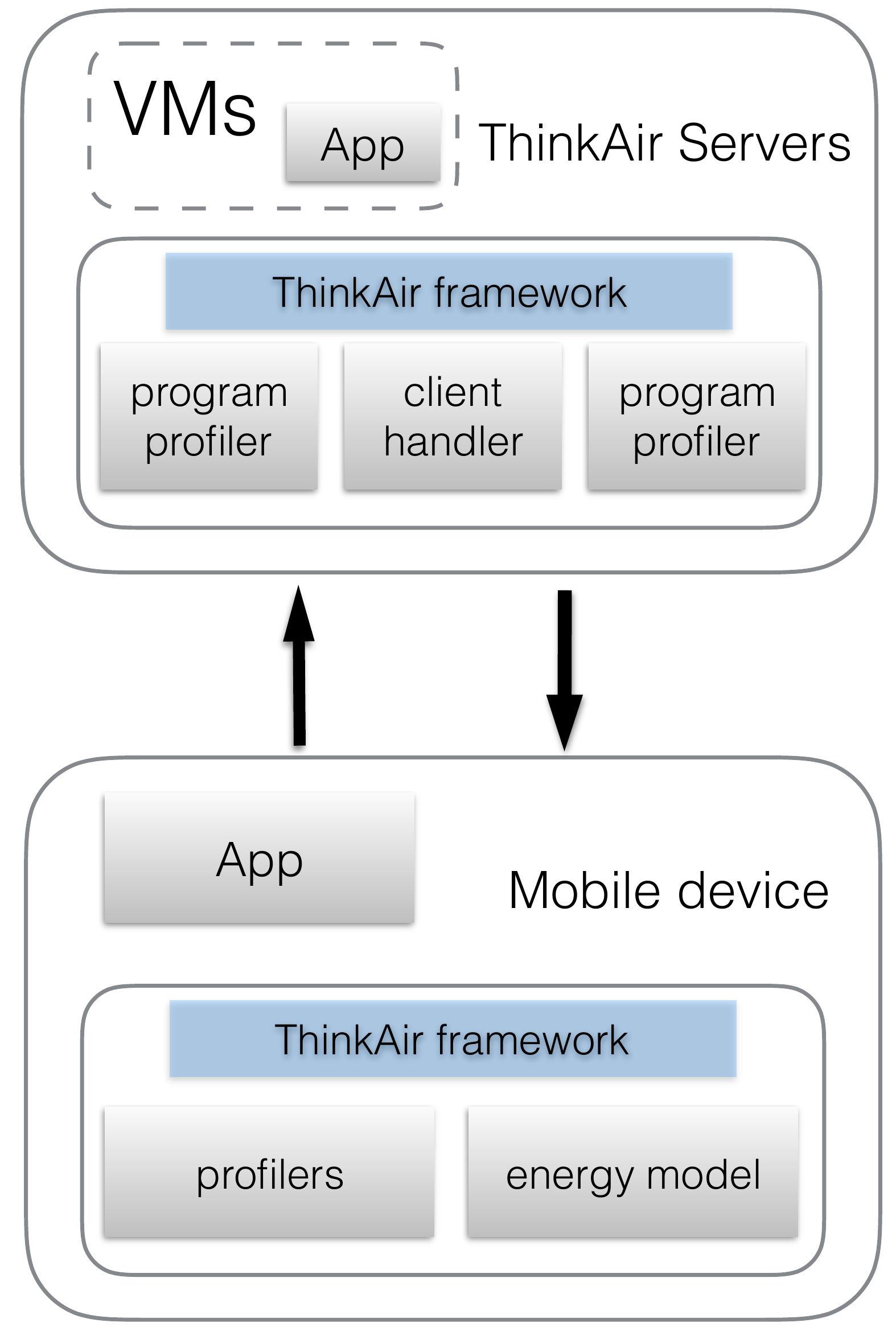}
\captionsetup{justification=centering}
\caption{Functional components in ThinkAir framework.}
\label{fig:thinkair}
\end{figure}

Fig.~\ref{fig:thinkair} shows an overall ThinkAir system framework and components. In the architecture of ThinkAir, the client handler and the profilers with rich resource are informed of the possibility of performing code migration. The client handler is deployed on the cloud servers to execute the tasks that require multiple VM clones in parallel. Moreover, the handler manages communication protocol, network connection, receiving and executing offloading code, and return results which are sent to the profiler for future offloading decision. The profilers consist of three modules: hardware, software and network. Specifically, the hardware profiler monitors the data of CPU, WiFi and 3G, which are also integrated into an energy estimation model. The software profiler records the data reported during application execution such as running time, CPU efficiency and memory usage, wherever in local or on cloud. The network profilers combine both intent and instrumentation profiling including bandwidth, response time and data of network interfaces. By utilizing the dynamic information collected with the monitoring processes within profilers, the system uses an energy estimation model so that the client handler can make efficient offloading decisions. 


\smallskip
\subsubsection{Cloudlet}
In contrast to ThinkAir, the Cloudlet is located at the closest access point which is the next hop to be connected to the Internet. The concept of cloudlet was initially proposed by \textit{Satyanarayanan}~\cite{Ma2}. He introduced the cloudlets as resource rich servers at the edge of the network located nearby WiFi access points. The mobile user could rapidly initiate VMs on the edge servers to offload its resource intensive computation. The main design goal of a cloudlet was to reduce application latencies compared to a case in which mobile devices are left on their own.

Despite the significant technology advancement provided by such a cloudlet system, \textit{Verbelen et al.}~\cite{Tim1} pointed out two drawbacks of this VM based cloudlet strategy. First, the cloudlets are provided by Internet Service Providers in their LAN (local area network) where cloudlets may have reduced range of operation and their configurations may fail to meet the execution requirements of some applications. Second, VM based cloudlets run the whole application offloaded in a single VM. However, the resources on the cloudlet are limited. When multiple applications are required to run simultaneously on the cloudlet, the cloudlet service will be impacted, e.g., some user requests could be declined. To overcome the above two limitations, authors in~\cite{Tim1} further proposed an elastic architecture of cloudlets. Their architecture not only provides original cloudlet servers at the edge but also organize ad-hoc clouds that involve other devices with available resource in the same LAN.

To deploy these new cloudlets, mobile applications are divided into different components that can be transmitted to other devices or basic cloudlet servers managed by a Cloudlet Agent (CA) running on a powerful server within the ad-hoc network. Such a component-level framework allows users to join and leave the cloudlet at runtime without severely impacting application performance. Each available device is regarded as single node that carries on a Node Agent (NA) and multiple nodes located in the physical proximity form a cloudlet which is monitored and controlled by the CA. When the offloading request appears, the NAs will estimate the execution environment and share the information with CAs. Based on the global view of resources on the nodes, CA can form a globally optimal solution. When nodes enter or quit the service coverage of the current cloudlet, the CA would perform calculation again and decide whether to migrate again some service components.

An augmented reality application is used to evaluate the cloudlet performance by splitting the application into five components: video source, renderer, tracker, mapper, relocalizer and object recognizer. The experiment shows how such flexible cloudlets indeed boost application performance. However, we must note that this kind of cloudlet needs to tackle execution scheduling of components carefully. The data synchronization among many of mobile devices and cloudlet servers can affect the global performance. In these cases, one cloudlet handles computation and data transmission for multiple applications. Meanwhile, an application may also split workload to multiple cloudlets. Given the complexity of offloading and high accuracy of synchronization, application developers and algorithm designers may face additional challenges.

\section{Offloading Balance - Online or Offline} \label{sec:balance}
\begin{figure}
\centering
\includegraphics[width=3.5in, height=2.3in]{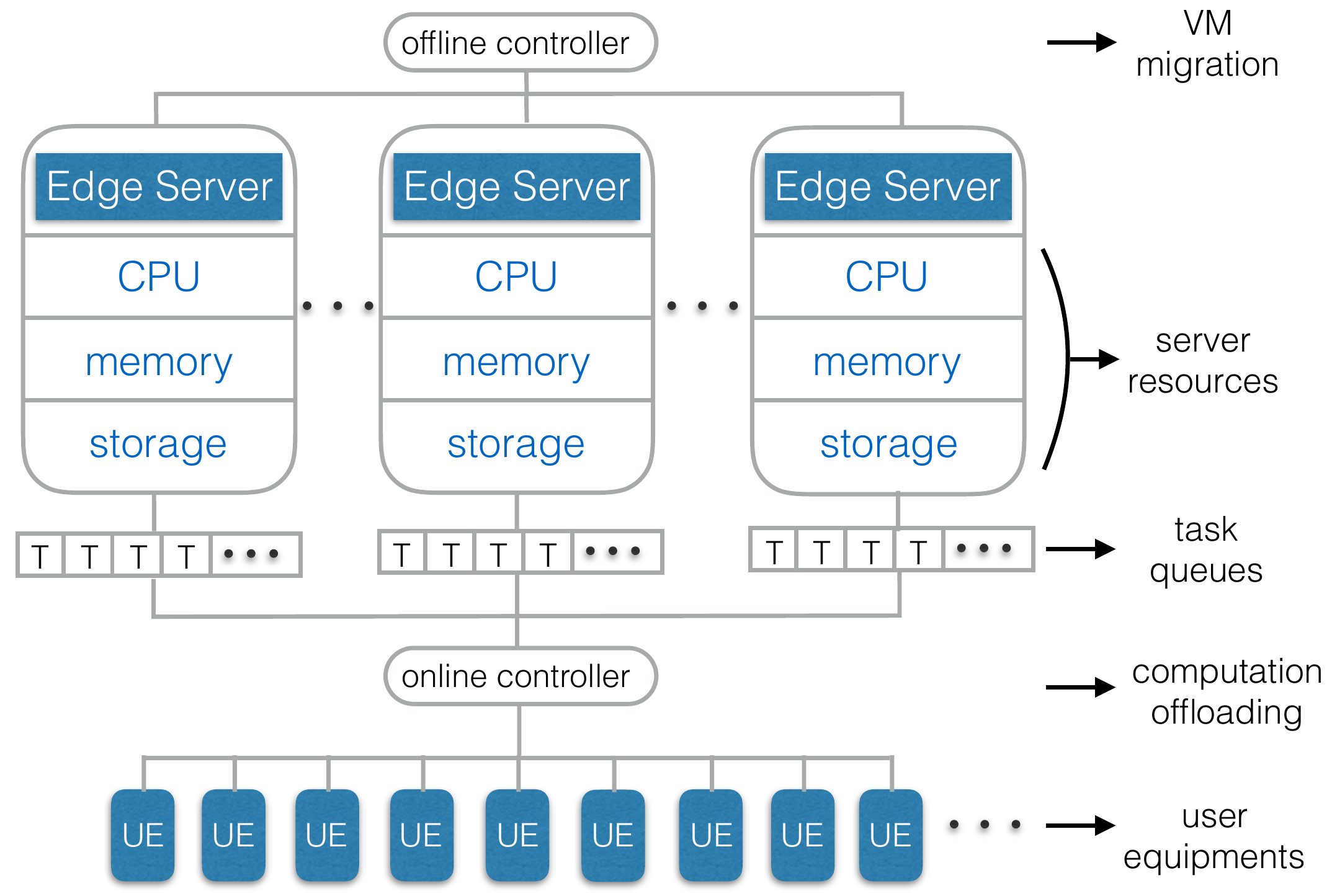}
\captionsetup{justification=centering}
\caption{The framework of balancing control system.}
\label{fig:balancing_system}
\end{figure}

In this section, we focus on offloading balance on solutions for distributed edge clouds despite their location. In some cases, the edge cloud may not have enough available resources making it hard to meet the Service Level Agreements (SLA) of the application; this means that not all user or application requests may be accepted. In these cases, the uncertainty on user request satisfaction and the weakness of offloading strategies may lead to unbalanced loads among cloud servers, where some parts of them may work with low loads while others are almost fully utilized. Under these scenarios, we classify the literature into two types of algorithms: online and offline. Fig.~\ref{fig:balancing_system} presents the framework of the balanced offloading system and the relationship among its functional components. An online controller distributes the tasks from users to edge servers and an offline controller migrates the workload among edge servers according to the resource usage. Next, we discuss a few relevant work based on such system.

\subsection{Online Balancing}
Online balancing is a pre-processing or run-time method that ensures that the workloads are distributed to the appropriate servers when the user requests arrive at cloud in real time. Online algorithms consider both user configuration requirements and current server's available capacity without any knowledge of future resource requests.

\smallskip
\subsubsection{Online-OBO and Online-Batch}
Considering limited resources and processing abilities on the edge cloud, \textit{Xia et al.}~\cite{online1} proposed an online request admission algorithm to maximize the system throughput. The offloading environment is a simple system where mobile devices connect to the cloudlet through an access point. There are different types of resources on the cloudlet such as network bandwidth, computing power, data storage space and service time slots, where any type of resource is associated with an estimated cost value.
 
The authors first propose an abstract admission cost model to determine different contributions from different resources in a cloudlet with K types of resources. Then they propose and implement a set of algorithms that handle the online requests admissions on the cloudlet without knowing the future request arrival rate according to the situation of resources occupancies on the cloudlet. When a request arrives at the system, it will be rejected immediately if there is insufficient capacity for every requested resource. Otherwise, the system verifies if the cost is under a given feasible threshold and if yes, it admits the request. After completing each request, the allocation mapping is updated, which contains all the functional components of the system.


Given the above admission control mechanism, the throughput maximization problem is modeled using a reduction from the online K-dimensional bin packing problem, a reduction typically utilized to solve space management problems. Two use cases are discussed in the paper~\cite{online1}: one request arrival per time slot and multiple requests arrivals per time slot. The former use case is implemented with the idea discussed above, referred as Online-OBO. The latter use case adopts a greedy strategy to try to handle a group of requests based on the system cost model, referred as Online Batch. Once a request is rejected, it will be removed for further consideration in the current time slot. It is worth mentioning that these two algorithms dynamically adjust some parameters in their mathematical models to offer convenient configuration of server workload for service providers. The details of the Online-OBO algorithm are shown in the Algorithm~\ref{alg1}.

\begin{algorithm}
\caption{Online-OBO balance~\cite{online1}}
\label{alg1}
\begin{algorithmic}[1]
\REQUIRE a request $i$ at time $t$ $q_{i}(t)$, available amount of resources $H(t)$, threshold of unit addmission cost of type $n$ resource $T_{n}$ (their average value $T_{avg}$) , unit admission cost $C_{i}(t)$, where $q_{i}(t):= \left \{ q_{i,1}(t),..., q_{i,N}(t)\right \}$, $\forall 1\leq n\leq N$.
\ENSURE accept or deny request $q_{i}(t)$
\item[]
\FOR{each $q_{i,n}(t)$ in $q_{i}(t)$}
\IF{$q_{i,n}(t)>H_{n}(t)$}
\STATE deny request $q_{i}(t)$;
\STATE EXIT;
\ELSIF{$C_{i,n}(t)>T_{n}$}
\STATE deny request $q_{i}(t)$;
\STATE EXIT;
\ELSE
\STATE update $H_{n}(t)$;
\ENDIF
\ENDFOR
\IF{$C_{i}(t) \leq T_{avg}$}
\STATE update $H(t)$;
\RETURN accept $q_{i}(t)$;
\STATE deny request $q_{i}(t)$;
\STATE EXIT;
\ENDIF
\end{algorithmic}
\end{algorithm}

\smallskip
\subsubsection{Primal-dual Approach}
Compared to the above online allocation approach where the users offload their workload to a single cloudlet server, \textit{Hao et al.}~\cite{online2} proposed an algorithm for allocating VMs in a distributed cloud which consists of geographically diversified small data centers close to users. The algorithm takes users' specified constraints into consideration including server geographic locations, restrict on resource cost, and communication latency for achieving:

\textbf{a.} Balance between optimal revenue and cloud performance. From the aspect of revenue, the system attempt to accept user requests as much as possible in the constant service time. Meanwhile, the system must maintain good performance to satisfy the SLA.

\textbf{b.} Generate the optimal solution without information of future arriving requests. The algorithms focus on obtaining the best allocations for the current requests based on the existing cloud distribution.

\textbf{c.} Be flexible to handle different resource constraints such as VM location, VM service duration, Inter-VM distance and cost policy of the service provider.
\smallskip

Due to the complexity of resource constraints and performance goals, the offloading is converted into a NP-hard problem which could be solved by an approximate approach. Once a request is receieved, the system tries to solve the primal solution and its dual solution, where the former decides VM allocation and the later present the upper limit for the optimal allocation. The primal and dual problem could be represented and solved by linear programming equations. Overall, this algorithm is a generalized solution through a comprehensive NP-hard approximation, considering the limits of server resources in both central and edge cloud architecture.

\smallskip
\subsubsection{Stochastic Models}
Besides the above method to transfer the offloading problem to a bin packing problem, a stochastic model was proposed by \textit{Maguluri et al.} with the aim of finding the maximum system throughput under various theoretical or practical constraints. The model is stochastic since the authors assume that the user requests arrive by way of a stochastic process~\cite{online3}. The authors studied two popular algorithms, pointed out their disadvantages, and improved the MaxWeight method~\cite{MaxWeight}. Offloading is in the form of VMs employed in multiple servers where each VM contains various types of resources. 

The authors analyze several algorithms using a centralized allocation strategy in which all user requests are received and handled by a central scheduler. First, the Best-Fit method is proved not optimal with a case study. Second, they show how the current MaxWeight approach~\cite{MaxWeight} is not optimal either since it is only suitable for some ideal condition, where the offloaded workload can be migrated between cloud clusters without a high cost. However, the MaxWeight algorithm is shown to be costly in practice, since the task execution may be disrupted when the tasks are all allocated at the beginning of each time slot. Third, the author also present a non-preemptive MaxWeight algorithm is designed to allocate workload based on the current server capacity.

The paper also discusses an online algorithm in which every server has an individual queue for job requests to route workloads as soon as they arrive. When the global balancing is considered, the join-the-shortest-queue routing is utilized to cooperate with MaxWeight algorithms. The scheduler allocates requests arrived according to the updated queue information and VM configuration, as introduced in Algorithm~\ref{alg2}. When the arrival rates of requests and the number of servers increase, the information of queue length leads to considerable communication overhead. In this case, Power-of-two-choices method with myopic MaxWeight helps reduce the costly overhead when all servers are identical. Specifically, two servers are randomly sampled, and the user request will be allocated to the server whose queue has less delayed jobs. In contrast, the pick-and-compare scheduling randomly chooses a server and compare it with the one allocated in the last time slot. All the above algorithms provide us optimal ways for throughput under specific system requirements to keep online load balance.

\begin{algorithm}
\caption{Myopic MaxWeight Scheduling based on Stochastic Models~\cite{MaxWeight}}
\label{alg2}
\begin{algorithmic}
\REQUIRE real-time requests; task queues
\ENSURE stabilized queues; usage percentage of resources in servers
\item[]
\STATE \textbf{Join Shortest Queue}:
\STATE A job with resource type $m$ arrives in the system at time slot $t$, which is inserted in the shortest queue. The chosen server $i$ is with type $m$ jobs queue $Q_{mi}(t)$.
\STATE
\STATE \textbf{Myopic MaxWeight Scheduling}:
\STATE \textit{Step 1:} Generate a super slot with a set of $n$ continuous time slots, represented by $T$, where $T = nt$
\STATE \textit{Step 2:} If the request arrives at the start of a super slot $T$. The optimal configuration is $$C_{i}(t)\in arg\ max\sum_{m}Q_{mi}(t)N_{mi}, N\in \mathbf{N}_{i}$$  Otherwise, the configure is chosen $$C_{i}(t)\in arg\ max\sum_{m}Q_{mi}(t)N_{mi}, N\in \mathbf{N}_{i}+N_{i}(t^{-})$$ where $N_{mi}$ is number of VMs with resource type$m$ , $\mathbf{N}_{i}$ is the set of configuration in server $i$, $N_{i}(t^{-})$ is VMs hosted at the beginning of super slot $T$.
\STATE \textit{Step 3:} Update the queue length information. $$Q_{mi}(t+1)=Q_{mi}(t)+W_{mi}(t)-N_{mi}(t)$$ where $W_{mi}(t)$ is the requests at time $t$.
\end{algorithmic}
\end{algorithm}

\subsection{Offline Balancing} 

Offline algorithms aim at balancing over-utilized resources on the edge servers. Since different servers contain a variety of services, they may have different occupancy percentages for computation and communication resources. When a server is overloaded, load migration may be performed to avoid service disruptions and wasting of resources, i.e., a user request may be rejected because one type of the required resource cannot be satisfied.

\smallskip
\subsubsection{Resource Intensity Aware Load (RIAL)}
VMs are deployed on the physical machines (PMs) that have limited hardware resources. When one type of resources on the PM is close to be completely occupied, the new requests for VM initiation will be rejected even if the usage of other resources is at low level. Due to this dilemma, such unbalanced allocation leads to waste of computation and energy resources of PMs and the system cannot reach the optimal throughout performance. Therefore, the Resource Intensity Aware Load (RIAL) balancing algorithm is proposed to efficiently migrate VMs among cloud severs while ensuring low migration cost at the same time~\cite{RIAL}.

Considering resource intensity, we mean that the amount of resource type is demanded for the service. A program may ask for several VMs simultaneously to support different functions, leading to intensive communication between these VMs. RIAL dynamically allocates the offloaded workload to the edge servers based on their current usage of computing resources. On the other hand, the VMs which exchange data commonly will be deployed in the same server to avoid migration cost.Meanwhile, the migration among PMs also maintains minimum performance degradation. 

For reducing the possibility of overloading, RIAL periodically checks the resource usage on each PM, seek and migrate the VMs among PMs. Both the migrated VMs and the PMs as a destination are derived by the multi-criteria decision-making method (MCDM) which establishes decision matrix within all types of resources. The ideal VM to migrate owns the highest occupancy of one type of resource and lowest utilization rate of another, as well as least data transmission with other VMs.

MCDM calculates the Euclidean distance between each VM and PM based on the corresponding migration cost. The VM with shortest distance is selected. The detailed equation of Euclidean distance is below:
$$D=\sqrt{\sum_{k=1}^{K}[w_{ik}(x_{ijk}-r_{ik})]^2+(w_{t}T_{ij})^2}$$
where $K$ is types of resources, $w_{ik}$ is the weight of resource $k$ in PM i where weight represent the priority of migration, $x_{ijk}$ is the usage of resource $k$ of VM $j$ in PM $i$, $r_{ik}$ is the largest percentage of occupancy of resource $k$ in PM $i$, $w_{t}$ is the weight of data exchanging rate, $T_{ij}$ is the communication rate of VM $j$ with other VMs in PM $i$.

\smallskip
\subsubsection{Bandwidth Guaranteed Method}
From what we have discussed above, the algorithms considered scenarios where users continuously appear and require some (network or application) service at a specific time interval, for a specific task. However, offloading can be performed also during extended time periods, for example during an entire day . In this case, multiple edge servers could work cooperatively to serve customers in daytime, but the quantity of user requirements will dramatically decrease as the midnight approaches. When the load is low, the resources using the VMs are utilized inefficiently. Hence, the distributed active VMs can be migrated to one server while their original physical machines could be turned off to reduce total energy consumption. However, migration can also lead to new challenges. The VMs migration technology requires sufficient bandwidth to copy the current memory state to the destination server in order to initiate new VMs that resume the original service. At the destination server, existing VMs should keep the minimum bandwidth for the current users. Therefore, both the migration and the maintenance of current services share the same physical link. The bandwidth guaranteed method described in~\cite{offline2} aims at solving such bandwidth competition to migrate VMs in a shortest time while maintaining the minimum bandwidth for user traffic. The migration time is defined as: 
$$T=\frac{M}{B_{m}-W}$$
where $M$ is the size of memory used by VMs, $B_{m}$ is total network bandwidth. $W$ is the current occupied network traffic. When $B_{m}\leq W$, migration is impossible. When $B_{m}>W$, the migration time depends on $B_{m}-W$.

In the proposed method, the order of the VM migrations is also considered. When the available bandwidth for VM migrations is abundant, the VM that has a large amount of state changes is migrated. Similarly, when the amount of available bandwidth is limited, the VM with small amount of state changes is migrated. The bandwidth guaranteed method has very practical contribution in achieving reduction of electric power consumption of service provider, which may be widely deployed as an elastic scheme to perform cloud control automatically.

\section{Mobility of devices}\label{sec:mobility}

Mobility poses new challenges to the offloading algorithm designer. As a mobile user moves across different service areas, the device may leave the service coverage area of its original edge server. Such mobility will lead to two problems: First, we need to decide if the edge service should be migrated out of the original server to a new server to keep the communication efficient. The migration deciding factor needs to resolve a tradeoff between the cost of long distance communication and migration cost. Second, the network signal, e.g., in WiFi and 3G/4G/5G, may be affected by large objects data transfers, heterogeneous network environments, and the connection policies of smart devices especially in the overlapped service areas. Persistent connectivity is not guaranteed and the intermittent connection is possible. In this section, we will discuss three representative approaches that handle the mobility problems in different environments under the edge cloud infrastructure.

\subsection{Offloading in Two-Tiered Mobile Cloud}
To improve the performance while satisfying the SLA of mobile applications, \textit{Xia et al.}~\cite{Mob1} proposed a two-tiered mobile cloud architecture that contains both the edge clouds and center clouds. Even if the edge cloud has the advantages of low latency and high scalability, the capacity of edge cloud may run into over-utilized problem when too many users offload their workloads to the same edge cloud which could suffer from longer delay and heavier energy consumption (such peak-load situation may happen when the assemblies are held by big groups of people in public place). The proposed algorithm aims at offloading location-aware tasks of mobile applications to local or remote servers to ensure fairness of energy consumption that battery life of each mobile device is prolonged equally. In this case, each device should consume the same portion of energy regarding its total energy capacity.

The two-tiered architecture supports an opportunistically flexible approach called Alg-MBM to help each mobile device choose the appropriate cloud server. The Alg-MBM constructs a weighted bipartite graph to find a weighted maximum matching offloading destinations including remote data centers, local edge servers and even mobile device itself. It is worth noting that executing locally on the mobile device instead of offloading may be the best choice when the outside computation resources are heavily costly due to poor network conditions. The details of Alg-MBM algorithm are shown in Algorithm~\ref{alg3}. 
\begin{algorithm}
\caption{Location-aware Alg-MBM~\cite{Mob1}}
\label{alg3}
\begin{algorithmic}[1]
\REQUIRE user requests $r(t)$ that arrives at time slot $t$, a local cloudlet $C_{e}$, a remote cloud center $C_{c}$
\ENSURE  maximize the battery life of each mobile devices to extend its work time
\item[]
\STATE {Step 1:} Collect the requirements of the requests $r(t)$ including resource consumption $R(t)$ and device location $L(t)$.
\STATE {Step 2:} Generate a bipartite graph $G(t)=(R(t),L(t))$ for requests arrived at time slot $t$, in which the weight of edges represents the energy cost.
\STATE {Step 3:}  Find the optimal path with least edge weight in total to complete all the request at time $t$.
\STATE {Step 4:} Execute workload in mobile devices, or offload to $C_{e}$ or $C_{c}$ according to the path in the previous step.
\end{algorithmic}
\end{algorithm}

\subsection{Follow Me Cloud}

Follow Me Could (FMC) is a framework in which the mobile devices move across edge severs while the cloud service smoothly supports the user applications~\cite{Mob2}. Because of the unpredictability of user mobility, VMs migration as the key technology for continuous cloud services breaks the limitation of geography. However, there are unresolved technical issues since migrating VMs may have two restrictions: the latency of converting a VM to be ready for migration and the latency to transmit VM state over the network among edge servers. On the other hand, if the destination servers use different hypervisor with the original server or the bandwidth is not qualified, the service migration becomes more costly and may even be rejected.

An algorithm based on Markov Decision Process (MDP) is proposed to determine whether such migrations should be performed when the user device is at a given distance from the original server~\cite{Mob3}. The authors defined a Continuous Time MDP (CTMDP) that contains the state of user devices, their transition probabilities and cost information. Moreover, they propose a Decision Time MDP (DTMDP) based on CTMDP with limited state spaces , which is regarded as a one-dimension model (1-D). Such a MDP is a static way to derive the optimal offloading since the migration cost function is pre-defined. Moreover, the finite state space will leads to high response time in solving the MDP.

To overcome the limitation of a static time cost calculation, a new dynamic service migration method is proposed to solve the limitations of MDP by \textit{Wang et al.}~\cite{Mob4}. The authors considered a two-dimensional (2-D) mobility model that considers a MDP with arbitrarily larger state space. 2-D mobility means that the user moves in a 2-D space. Since both network topology and user mobility continuously changes, practically the cost function and transition probabilities of MDP may fluctuate frequently. Therefore, the MDP should be solved in a dynamic manner. The 2-D mobility algorithm can obtain an approximate solution by applying the distance-based MDP. Meanwhile, it decreases one order of magnitude of the overall complexity in each MDP iteration to improve time efficiency.

\begin{figure}
\centering
\begin{minipage}{.5\textwidth}
	\includegraphics[width=2.2in, height=1.5in]{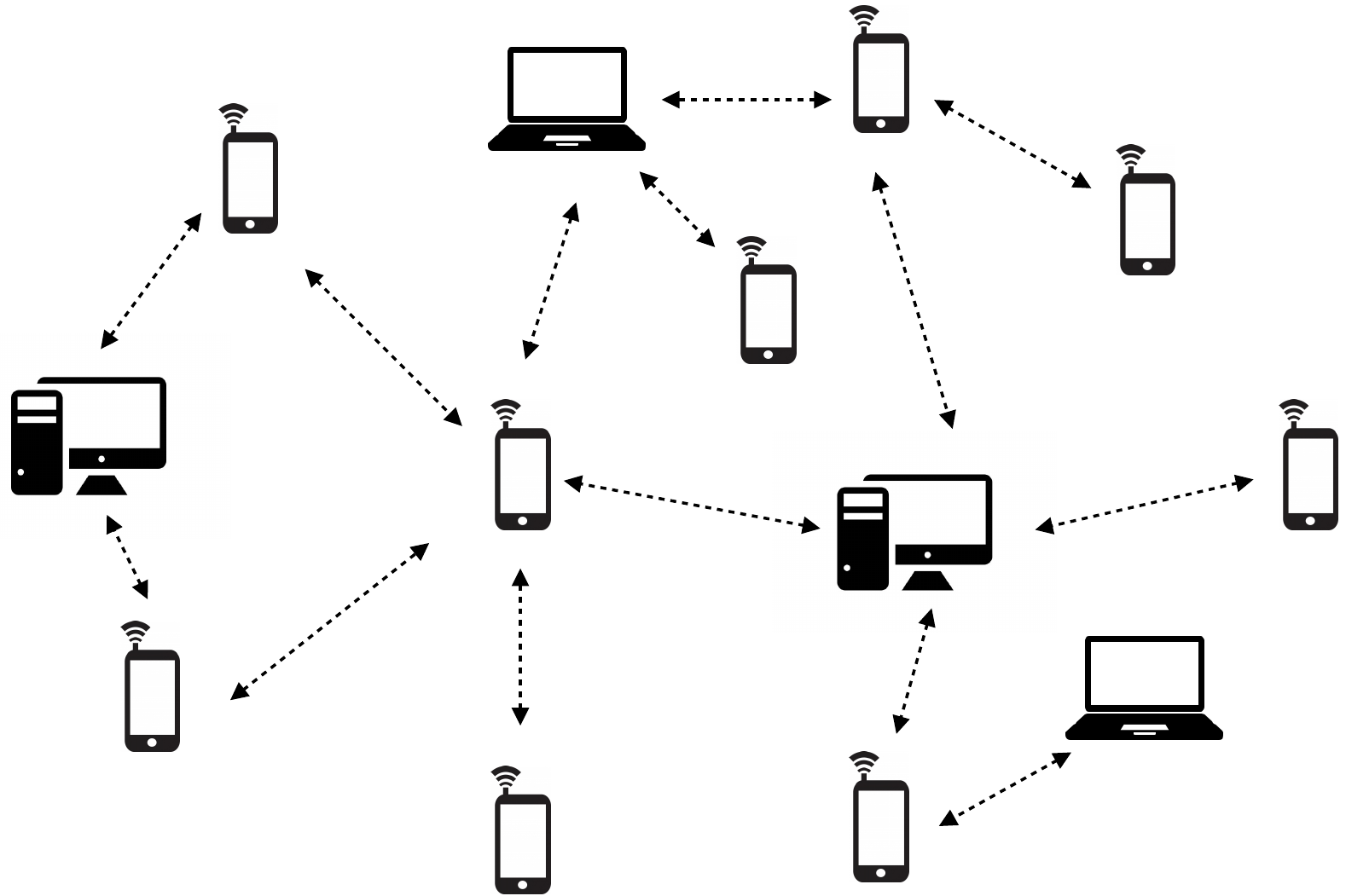}
	\captionsetup{justification=centering}
	\caption{Self-organized ad-hoc mobile cloud.}
	\label{fig:adhoc_cloud}
\end{minipage}%
\begin{minipage}{.5\textwidth}
	\includegraphics[width=2.6in, height=1.5in]{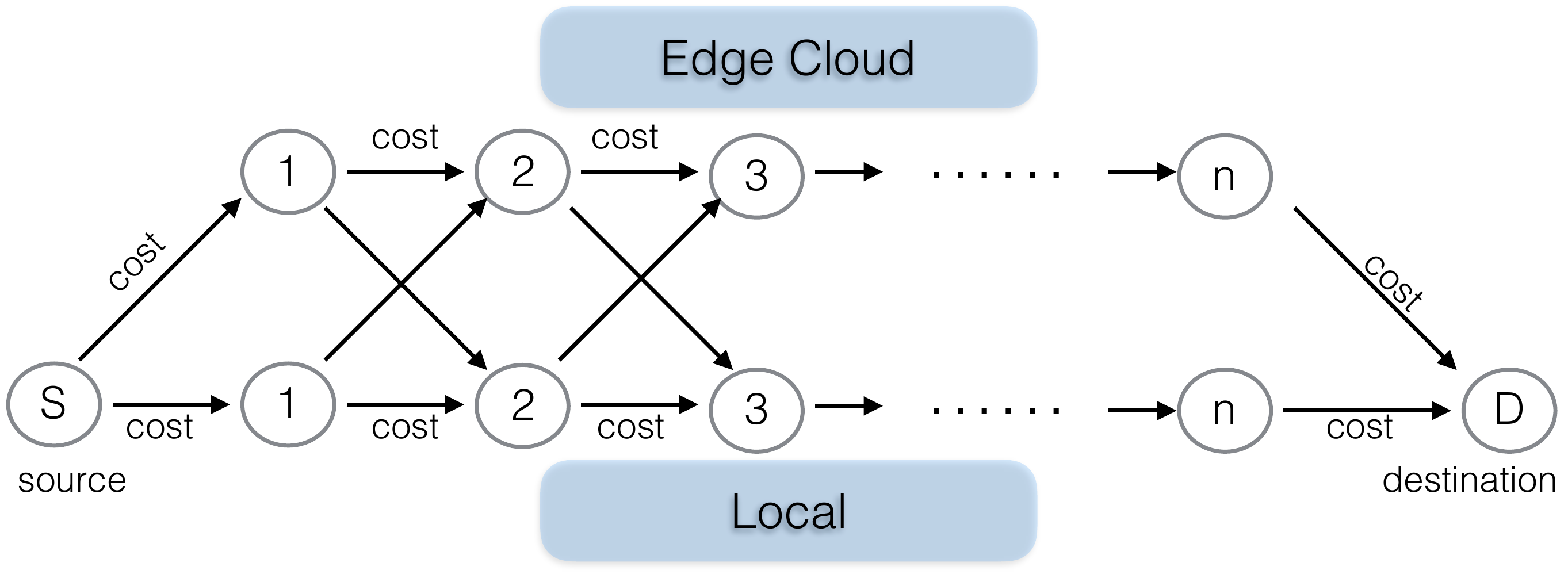}
	\captionsetup{justification=centering}
	\caption{Offloading choice control flow graph.}
	\label{fig:choice_control}
\end{minipage}
\end{figure}

\subsection{Ad Hoc Cloud Assisted Offloading}
Besides the two-tiered mobile cloud and FMC (Follow Me Cloud), mobile ad hoc networks are another important applications that motivates researchers to pursue higher resources utilization and deal with user mobility. When the intermittent connectivity happens, a type of ad hoc self-organized mobile cloudlets help mobile devices obtain close computation resources from other idle terminal devices including smartphones, laptops and desktop computers to form a self-organized mobile cloud, as illustrated in Fig.~\ref{fig:adhoc_cloud}. As discussed in the Section 2, Cloudlet provides such an ad hoc architecture which integrates ad hoc mobile device cloud with infrastructure-based edge servers~\cite{Tim1}~\cite{Tim2}.

Based on such a mobile cloud at network edge, an up-to-dated centralized task scheduling (CTS) algorithm was proposed by \textit{Wu et al.} to guarantee SLA and achieve energy consumption balance~\cite{Mob5}. As discussed in the algorithm, the execution of mobile application could be represented by a control flow graph which contains the computation components working in flow. A collaborative task execution scheme is implemented to determine where the components execute, local device or edge cloud, as illustrated in Fig.~\ref{fig:choice_control}. Aside from computation cost, the data transmission cost between edge and local is considered. 

Under such a collaborative scheme, offloading is more flexible. When the infrastructure-based cloudlet is unavailable to execute assigned work tasks, the centralized task scheduler starts to seek the available mobile devices resource in the certain range by qualification judgement based on the current usage of resource on the mobile cloud. The judgement process is formulated as a 0-1 integer linear programming problem and approximately solved by the greedy algorithm to get a solution and keep low complexity.


\section{offloading partition} \label{sec:offloadingPartition}

Under the premise that the offloading achieves low latency, researchers make their best endeavors to prolong the battery life. Since the increment of battery capacity cannot catch up the fast advance of program and application technologies like virtual reality and augment reality, it is impossible to run all the parts of such applications only on the mobile device. The division and organization of partitioned components are the foundations to design an offloading algorithm. Therefore, the algorithms of computation partitioning are further studied to determine which parts of the user application are offloaded and how they executed in order. The current partition strategies can be divided to into three aspects: static, dynamic, and a combination of static and dynamic.

\subsection{Static Partition}
In the early years of research on cloud offloading, research studies were proposed on offloading computation of mobile devices to a close powerful server in the same LAN through wireless connection. \textit{Li et al.}~\cite{Li1} proposed a static partition approach based on the cost graph generated by the data of computing time and data sharing at the level of procedure calls. The cost graph statically divides the application program into server tasks and mobile device tasks to minimize energy consumption. Moreover, the program execution follows the order of sequential control flow. The data shared between two tasks is sent by the push and pull method which guaranteed that the server and client continuously update the most recent data modifications.

Based on such a scheme, the cost graph contains computation and energy information during the whole execution of sequential tasks. Then a Branch-and-Bound algorithm defines the offloading problem by linear expression to calculate optimal solution. However, the worst-case complexity of Branch-and-Bound is unacceptably costly according to the cost graphs of some applications. In this case, a pruning heuristic method is proposed to reduce the calculation time by only focusing on the components with heavy workload. 

\begin{figure}
\centering
\includegraphics[width=3in, height=1.15in]{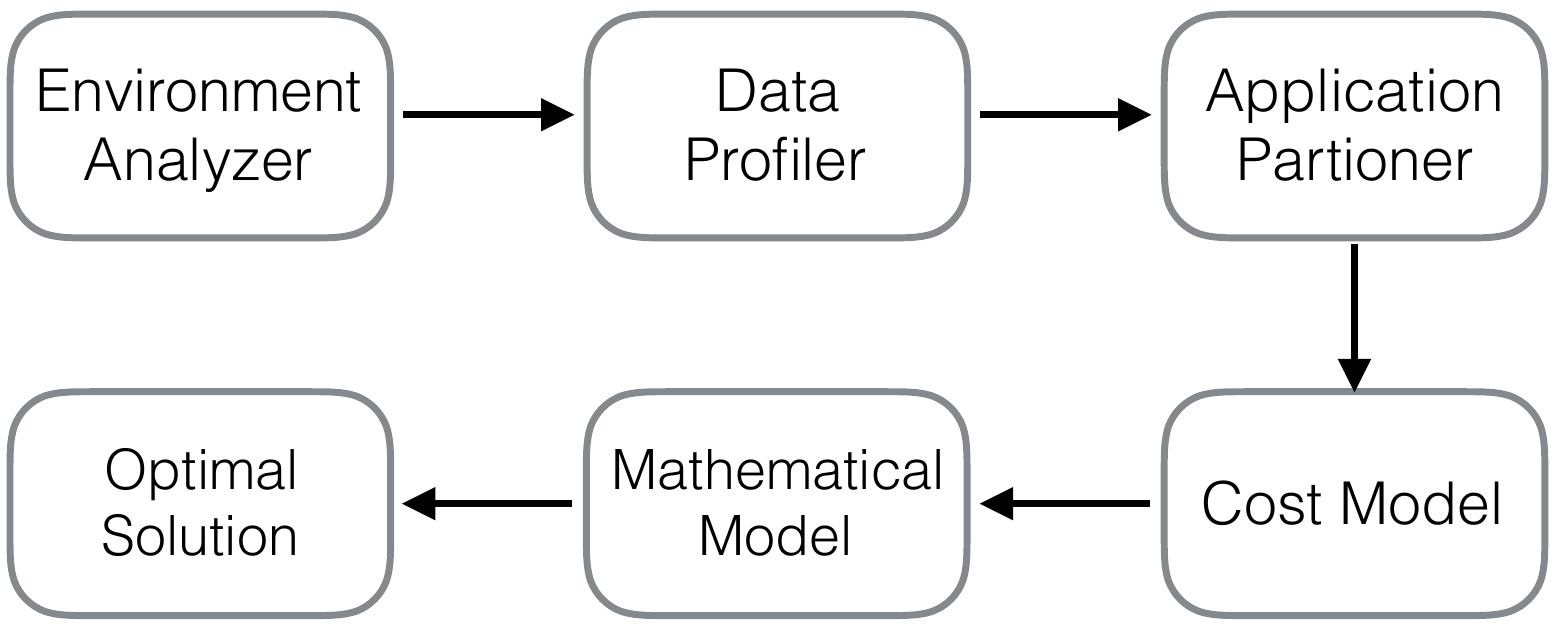}
\captionsetup{justification=centering}
\caption{Flow graph of offloading partition.}
\label{fig:offloading_partition}
\end{figure}

This scheme is static because all the profiled information is based on the intrinsic characteristics of the program which leads to only one optimal solution. Fig.~\ref{fig:offloading_partition} presents a flow graph used to the partitioning algorithms discussed above.

\subsection{Dynamic Partition}
While the static method considers all the parameters of the system and generates a globallyl optimal solution, the dynamic methods are more flexible as they also evaluate network and server states. 

The Dynamical Offloading Algorithm (DOA) proposed by \textit{Huang et al.}~\cite{dynamic1} focuses on achieving energy saving given the change of communication environment. Meanwhile, the interdependency of the partitioning application components should be considered because of the different execution latency constraints and data sharing cost with each other. In~\cite{dynamic1}, \textit{Huang et al.} proposed a DOA to offload partitioned components with the change of wireless connection. They also created a cost graph where the vertices present application modules and the directed edges are data sizes from the source vertex to the destination vertex. In this case, the data transmission rates based on wireless environment take dynamic effect on the decision of offloading, locally or remotely. The energy consumption and the total application execution time are formulated as a Lyapunov optimization problem which introduces a control parameter to take a tradeoff between energy and latency.

Besides the uncertainty of network connectivity, the dynamic partitioning for saving energy should take more heterogeneous factors into account~\cite{dynamic2}. Such factors include device operating system, network types, cloud availability and user workload. For example, an image matching program contains three stages: image feature generation, similarity calculation against a database, and classification. The content complexity of images are various and the size of the matching database is changing so that the image processing may take a longer or shorter time during different stages. Given such an application, its device platform can be smart phones, tablets or laptops with a wide variety of CPU, memory and storage resources. The network is assumed to be a 3G/4G cellular network or a WiFi with different bandwidth. The cloud providers are assumed to have different prices and performance for their services, while the workload is dynamic at different stages. Additionally, another algorithm is proposed~\cite{dynamic3} to systematically discusses the factors affecting the performance of real-time video based applications in dynamic wireless network condition. In~\cite{dynamic4}, an algorithm based on a dynamic programming with hamming distance terminations is proposed, which mainly focuses on available network bandwidth.

\subsection{Combination of Static and Dynamic Partition}
We could also combine the static and dynamic approaches to build a model for optimal offloading partition. \textit{Giurgiu et al.}~\cite{Giu1} propose a cost flow graph which is established based on the functional units and interdependency degree in terms of resources consumption such as data sharing, code size and memory cost, similar to Fig.~\ref{fig:offloading_partition}.

Two partitioning algorithms in ~\cite{Giu1}, are proposed to derive static and dynamic optimization separately: ALL and K-step. ALL determines the best partitioning by evaluating all offline information of application and network. In addition, K-step performs partitioning of applications in real time when the mobile devices request services from cloud servers with their specific requirements. The algorithm starts estimating one node at a time by combining depth-first and breadth-first method till the last node. Compared to ALL, K-step is faster because it considers only a reduced set of configuration and less accurate. If there is a new configuration on nodes that offers a better solution, a new local optimum will be updated.

\section{Partitioning Granularity} \label{sec:granularity}

\begin{figure}
\centering
\includegraphics[width=1.7in, height=1.6in]{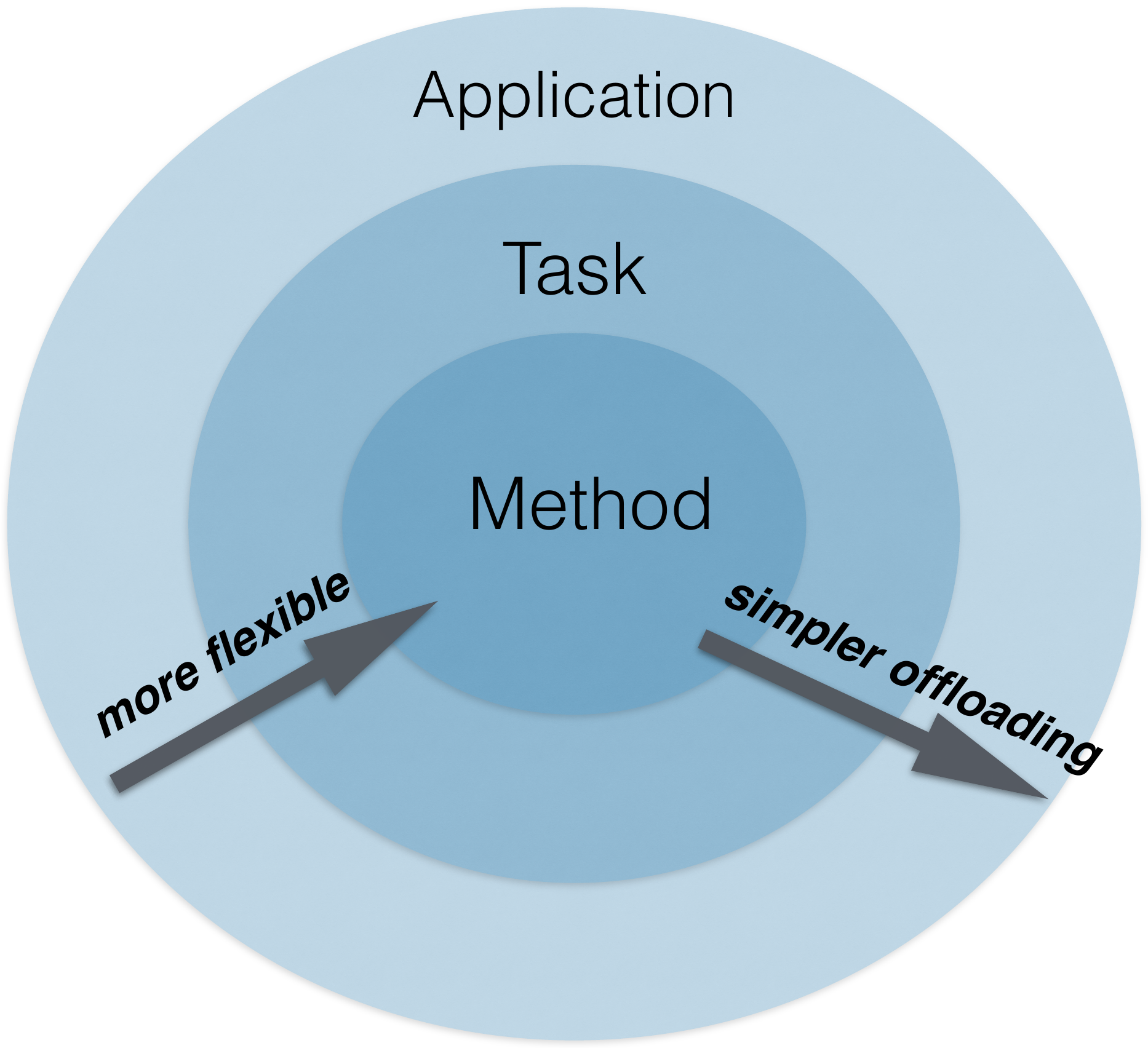}
\captionsetup{justification=centering}
\caption{Inclusion relationship of granularities.}
\label{fig:granularities}
\end{figure}

Before offloading computation to the edge servers, we must also consider the rational size of components that could run remotely. Given that different applications consist of the customized functional components designed by their developers, the partition granularity is a significant factor to improve the global execution performance. The granularity of partitioning is defined as the different sizes of offloading components. In this section, we classify the granularity of partitioning into three levels (from the biggest scale to smallest): Application, Task Module and Method. Fig.~\ref{fig:granularities} offers an overview of their relationship. The more fine-grained granularity, the more flexible and complex the offloading system is. Meanwhile, the comparison of advantages and disadvantages are summarized in Table ~\ref{tab:one}.

\begin{table}%
\small
\caption{Offloading granularity comparison}
\label{tab:one}
\begin{minipage}{\columnwidth}
\begin{center}
\begin{tabular}{l|l|l}
 \toprule
 \thead{Granularity}  & \thead{Advantages} & \thead{Disadvantages}\\ \hline
Application & \vtop{\hbox{\strut 1.thin workload on mobile devices}\hbox{\strut 2.easy VM configuration on servers}} & relatively long time of VM initialization\\  \hline
Task & \vtop{\hbox{\strut 1.offer flexibility to developers}\hbox{\strut 2.less synchronization work}} & low reuse posibility of execution environment\\  \hline
Method & wider authority to developers & \vtop{\hbox{\strut 1.high complexity to achieve optimal offloading}\hbox{\strut 2.harder data synchronization}\hbox{\strut 3.more fragmented user requests}} \\
  \bottomrule
\end{tabular}
\end{center}
\end{minipage}
\end{table}%

\subsection{Application Level}
The computation components at the application-level contain the whole functionalities of software; this is the case, for example, for face recognition and voice translation applications. The partitioning algorithms at the application level impose a very thin workload to the mobile devices. The corresponding software has already existed on the server that only needs to configure initialization data. Virtual machines (VMs), whose images are preinstalled on the servers, are most commonly utilized to meet the requirements at this granularity level. One of the advantages of application level partitioning is an easier system configuration on the server side where the initiated VMs can be simply removed and new clean VMs can be ready for the next service time slot. In this case, the mobile devices do not need to upload any functional parts because the entire computation is executed on servers. Meanwhile, the less-intensive tasks, such as user interface and system data management, run on mobiles devices without high energy cost. The Cloudlets~\cite{Tim1}~\cite{Ma2} we introduced in the previous parts adopt this application-level partitioning for offloading.

\subsection{Task Module Level}
The offloaded parts in the task module level are the application elements whose responsibilities are separated in a sequential or parallel order. A typical example is the face recognizing program which sequentially executes face detection, face verification and face identification. As discussed in~\cite{dynamic1} and~\cite{Task1}, the cost model for task control flow, which is derived from the specific system features, displays the tasks partitioning at this granularity level. Such task modules generally execute in the containers of code running environment such as Java virtual machine (Java Run Environment), .Net framework and self-built running platform. Task module level offloading is relatively more flexible than application level partitioning since the developer is allowed to make decisions to maximize performance using the edge cloud. Moreover, every task module is relatively enclosed which receives input data from the previous stage and return its state to the next stage where there is not much synchronization work between the mobile devices and edge cloud. However, task module as a medium granularity form puts forward more rigorous demand to the cloud servers where the running environment faces more diverse requests from users. The environment configuration from user requests may ask for different classes of libraries which lead to the lower possibility of container reuse. On the other hand, the programming languages available to developers are also relatively limited by running platforms.


\subsection{Method Level}
Method is in a lower level than task module for partitioning, which can also be presented in form of functions as a code fragment. MAUI~\cite{MAUI} and ThinkAir~\cite{ThinkAir} require the developers to manually or semi-automatically annotate the methods as offloading permitted. The advantage of method level partitioning is that the developers have wider authority to improve their applications. However, such low-level granularity brings several challenges to achieve offloading optimization. First, the high complexity of obtaining optimal solution may take longer time because of a large number of methods. Second, the data synchronization between local devices and remote servers is harder to guarantee while they should share same execution results. A synchronizing scheme runs periodically or in real time to collect and update the method data. Third, the service deployed on the server will handle more fragmented requests which need a robust identification mechanism to distinguish their source applications.

\section{Discussion and Perspectives}\label{sec:discussions}

After presenting the above research work from five perspectives, in this section, we present some analysis and discussions on the mathematical models of existing algorithms, potential challenges, and technological trends for future offloading on edge cloud.

\begin{table}%
\small
\caption{Characteristics of offloading algorithms}
\label{tab:two}
\begin{minipage}{\columnwidth}
\begin{center}
\begin{adjustbox}{angle = 90}
\begin{tabular}{l|l|l|l|l|l|l}
  \toprule
  \thead{Algorithms} &  \thead{Destination} &  \thead{Balance} &  \thead{Mobility of Device} &  \thead{Partition} &  \thead{Granularity} &  \thead{Mathematic Model}\\ \hline
MAUI[4] & single server & online & \vtop{\hbox{\strut    simply restart latest}\hbox{\strut disconnected offloading}} &dynamic &  method & linear programming \\  \hline
CloneCloud[5] & single server & online & NS (not specified) &dynamic &  method & linear programming\\  \hline
ThinkAir[6] & multiple servers & online & NS &dynamic &  method & linear programming\\  \hline
Cloudlet[7] & multiple servers & online & NS &dynamic &  application & linear programming\\  \hline
\vtop{\hbox{\strut Online-OBO and}\hbox{\strut Online-Batch[9]}} & single server & online & NS & NS &  application & K-dimension bin packing \\  \hline
Primal-dual approach[10] & multiple servers & online & NS & NS &  application & NP-hard primal-dual approach\\  \hline
Myopic Maxweight[11] & multiple servers & online & NS & NS &  application & \vtop{\hbox{\strut myopic MaxWeight algorithms}\hbox{\strut with various routing policies}}\\  \hline
RIAL[12] & multiple servers & offline & NS & NS &  application & \vtop{\hbox{\strut multi-criteria decision }\hbox{\strut making method}}\\  \hline
Bandwidth Guaranteed[13] & single server & offline & NS & NS & application & self-designed model\\  \hline
Alg-MBM[14] & single server & online & \vtop{\hbox{\strut center cloud and edge cloud}\hbox{\strut work cooperatively in the}\hbox{\strut two-tiered architecture}} &dynamic &  task & \vtop{\hbox{\strut weighted maximum matching}\hbox{\strut problem in the bipartite graph}} \\  \hline
1D MDP[16] & multiple servers & online & \vtop{\hbox{\strut cooperation among }\hbox{\strut edge servers}} & static &  NS & Markov Decision Process \\  \hline
2D MDP[17] & multiple servers & online & \vtop{\hbox{\strut cooperation among }\hbox{\strut edge servers}} & static &  NS & Markov Decision Process \\  \hline
Ad Hoc assisted CTS[19] & multiple servers & online & \vtop{\hbox{\strut cloudlets and ad hoc }\hbox{\strut mobile devices }} & dynamic &  task & 0-1linear programming \\  \hline
Static partition scheme[20] & single server & online & NS & static &  task & \vtop{\hbox{\strut branch and bound}\hbox{\strut + pruning heuristic}}\\  \hline
ALL and K-Step[21] & single server & online & NS & \vtop{\hbox{\strut static +}\hbox{\strut dynamic}} & task & self-designed model\\  \hline
DOA[22] & single server & NS & \vtop{\hbox{\strut center cloud and edge cloud}\hbox{\strut work cooperatively in the}\hbox{\strut two-tiered architecture}} &dynamic &  task & Lyapunov optimization \\  \hline
\vtop{\hbox{\strut Joint optimization}\hbox{\strut algorithm[23]}} & multiple servers & NS & NS & dynamic & task & convex optimization \\

  \bottomrule
\end{tabular}
\end{adjustbox}
\end{center}
\end{minipage}
\end{table}%


\smallskip
\noindent
\textbf{Mathematical Models: }
We further summarize and compare the algorithms discussed in this paper into Table ~\ref{tab:two}. Besides the proposed five categories, we classify the related work also based on the mathematical models and optimization methods utilized. Examples of such methods include 0-1 integer linear programming problem, K-dimensional bin parking, Markov decision process and Lyapunov optimization. Since most of these optimization models attempt to solve an NP-hard problem, the approximation solutions with higher performance and lower complexity are designed and evaluated in many cases. When researchers try to include more factors or constraints in their offloading algorithms to meet specific performance requirements, the algorithms need to be improved to be more flexible. For example, a hierarchical edge cloud architecture was proposed to solve the offloading problem during the peak hours~\cite{Diss1}. Dynamic voltage scaling technique was implemented to vary the energy supplement of mobile devices based on the computation loads~\cite{Diss2}. To address the potential requirements of future applications, the existing algorithms should be adjusted and new approaches need to be explored.

\smallskip
\noindent
\textbf{EC Offloading for Future IoT Environment: }
The Internet of Things is estimated to bring the next major economic and societal revolution by turning billions of independent electric devices into an enormous interconnected community where the data shares more frequent and fast than ever before.~\cite{Book1} According to the forecast of IHS and Mckinsey \cite{Web1}~\cite{Web2}, the devices connected to IoT network will increase from 15.4 billion in 2015 to 30.7 billion in 2020 We can imagine that the future society will be boosted by seamless intelligent cooperation among smart devices, and the creation of smart-x applications such as smart health, smart home, smart city, smart energy, smart transport, smart farming and food security. 

Under the background of IoT, edge computing can potentially support significant progress to solve problems including communication latency, energy saving, user mobility, the variety of personalized applications, support of real-time applications and network heterogeneity. However, the research in this aspect is not mature enough yet to accommodate various IoT standards, and the current work still cover only a limited range of application scenarios. Many smart-x applications have not yet adopted corresponding customized cloud computing models very effectively. Therefore, the future research can focus on implementing a specific class of IoT systems and related algorithms.

\smallskip
\noindent
\textbf{EC Offloading for Big Data: }
With the development of Big Data technologies, media transmission and targeting delivery of customized content can improve the service efficiency and accuracy for mobile users. In conjunction with edge cloud computing, applying Big Data techniques, researchers can improve the performance of data processing such as collecting, capturing, analyzing, searching, exchanging, transmiting, and protecting. e.g, a cloud platform for the medical education to measure patients' personal big data~\cite{ali2017iotflip}. By offloading heavy workloads to edge servers, both computation and communication latency are guaranteed to extract valuable information from data. Facilitated by edge cloud, the application of Big Data could be conveniently accessed by terminal users. In this case, how to maximize the advantages of EC by effective offloading approaches to boost Big Data technology is still unexploited field.


\smallskip
\noindent
\textbf{EC Offloading for 5G: }
In the future, 5G will bring us broader network bandwidth and greater convenience of device connectivity which allows a larger number of mobile users per area unit. The mobile devices are also provided with high availability and speeds of network access. However, the spectrum resource is limited which could bear heavier burden in the 5G era. It may casue the increment of cost when cloud resources are accessed through the 5G network. Currently, there are some related researches of offloading in such situation. For example, a time-adaptive heuristic algorithm with multiple radio access technology (Multi-RAT) is proposed~\cite{Diss3}.  A distributed computation offloading algorithm solves the offloading decision-making problem in the wireless network with multi-channel in order to avoid mutual interference~\cite{liu2016multi}. To improve the performance of offloading in 5G network, EC can play an important role to enhance its upper bound.

\smallskip
\noindent
\textbf{Security Problems with EC Offloading: }
Security is another serious challenge for the future edge cloud service. First, data offloaded from mobile devices are exposed to the multiple threats: malicious eavesdropping, changing contents of data packets, destroying channel connectivity and uploading destination. The confidentiality, access controllability and integrity can not easily be guaranteed. Second, the orchestration of resources in the edge servers may be disturbed and the reliability and dependability of edge service can be compromised. In this case, the initialization and migration of VMs may not accurately match application requirements. Third, the IoT function visualization needs data security to align horizontal and vertical input information by utilizing SDN (Software-Defined Networking) and NFV (Network Functions Virtualization) technologies~\cite{security1}. Moreover, the data synchronization of an application should not be disrupted even if a single communication access bears multiple applications.

\section{Conclusions}\label{sec:conclusion}
In this paper, we collected and investigated the key issues, methods, and various state-of-the-art efforts related to the offloading problem in the edge cloud framework. We adopted a new characterizing model to study the whole process of offloading from mobile devices to the edge cloud, which consists of the basic categorizing criteria of offloading destination, load balance, mobility, partitioning and granularity. The overall goal of offloading is to achieve low latency and better energy efficiency at each step of computation offloading. An integrated offloading system of edge cloud should be a well-balanced combination of these five perspectives to properly solve offloading issues. The factors of algorithms such as environment constraints, cost models, user configuration and mathematical principles were discussed in detail. We endeavored in drawing an overall ``big picture'' for the existing efforts. Embracing the future network development, we plan to continuously explore emerging technologies and creative ideas that improve the offloading performance.

\begin{acks}
The work is supported in part by National Security Agency (NSA) under grants No.: H98230-17-1-0393 and H98230-17-1-0352, National Aeronautics and Space Administration (NASA) EPSCoR Missouri RID research grant under No.: NNX15AK38A, National Science Foundation (NSF) award CNS-1647084, and a University of Missouri System Research Board (UMRB) award.

\end{acks}

\bibliographystyle{ACM-Reference-Format}
\bibliography{references}

\end{document}